# Improving a pavement-watering method on the basis of pavement surface temperature measurements

Martin HENDEL[1,2,3*], Morgane COLOMBERT[2], Youssef DIAB[2,4], Laurent ROYON[3]

[1]Paris City Hall, Water and Sanitation Department, F-75014, Paris, France
[2]Université Paris-Est, EIVP, F-75019, Paris, France
[3]Univ Paris Diderot, Sorbonne Paris Cité, MSC, UMR 7057, CNRS, F-75014, Paris, France
[4]Université Paris-Est, LEESU-GU, UMR MA 120, F-77420, Champs-sur-Marne, France
[*](corresponding author: martin.hendel@paris.fr)

**Abstract:** Pavement-watering has been studied since the 1990's and is currently considered a promising tool for urban heat island reduction and climate change adaptation. However, possible future water resource availability problems require that water consumption be optimized. Although pavement heat flux can be studied to improve pavement-watering methods (frequency and water consumption), these measurements are costly and require invasive construction work to install appropriate sensors in a dense urban environment. Therefore, we analyzed infrared camera measurements of pavement surface temperatures in search of alternative information relevant to this goal. Firstly, surface temperature reductions of up to 4°C during shading and 13°C during insolation were found. Secondly, the infrared camera successfully detected temperature spikes indicative of surface drying and can therefore be used to optimize the watering frequency. Measurements made every 5 minutes or less are recommended to minimize relevant data loss. Finally, if the water retaining capacity of the studied pavement is known, optimization of total water consumption is possible on the sole basis of surface temperature measurements.

## 1. Introduction

Pavement-watering is only a recent topic in French cities such as Paris and Lyons [1]–[4], while it has been studied as a method for cooling urban spaces in Japan since the 1990's [5]–[10]. This technique is viewed as a means of reducing urban heat island (UHI) intensity, with reported air temperature reductions ranging from 0.4°C at 2-m [2] to 4°C at 0.9-m [7]. In France and especially Paris, the predicted increases in heat wave intensity and frequency due to climate change [11] and the high sensitivity of dense cities to such episodes [12]-[13] have focused research efforts on the search for appropriate adaptation tools. In parallel to techniques such as developing green spaces and improving urban energy efficiency, pavement-watering is seen as one of these potential tools.

As climate change is also expected to change the region's seasonal precipitation distribution [14], water use optimization of this technique is crucial. The City of Paris has taken an interest in pavement-watering and numerical and experimental studies of the method have been conducted over the last few years [1]–[3]. Previous work by the authors based on these experiments revealed that pavement heat flux and solar irradiance measurements could be used to optimize a pavement-watering technique applied to an approximately N-S street with a canyon aspect ratio equal to one (H/W=1) [3]. It was found that the optimum watering method consisted in watering the exact water-holding capacity of the pavement at the lowest frequency which avoids surface drying between watering cycles. Therefore, if the water-holding capacity of a given surface is known, then the optimization of pavement-watering rests fully on determining the lowest watering frequency necessary for the surface to remain wet. In the case of the studied street configuration and pavement materials, sprinkling 0.16-0.20 mm (equivalent to L/m²) every 60 minutes during shading and every 30 minutes during insolation was recommended. Conducted continuously from 06:30 until 18:30, this would result in the daily use of less than 3.2 mm/d. Generalized to all of Paris' 2,550 ha of street surfaces [1], this would amount to approximately 82,000 m$^3$/d or 36 L per day per capita, i.e. equivalent to one shower per person.

Unfortunately, streets within the same city often have different configurations or use different materials, preventing the generalization of conclusions drawn from a single site. It is therefore recommended to study several characteristic streets before a city-wide strategy can be developed. However, installing a heat flux sensor in combination with solar instruments over long periods of time is an expensive and invasive procedure and requires close cooperation with the relevant city services. It is therefore difficult to install large numbers of these sensors in a dense urban environment for this purpose.

In order to overcome these issues, we propose an alternative measurement method consisting of a fixed infrared camera recording radiometric data from selected areas of pavement at regular intervals. These instruments require very little effort to install compared to pavement heat flux or solar instruments and can therefore be installed at several urban locations more easily. Infrared temperature data will be analyzed in a similar fashion to that conducted for pavement heat flux [3] in the same hopes of minimizing the number of watering cycles while keeping the pavement surface wet. The effects of pavement-watering on surface temperatures will be studied in the process. Once the optimal watering frequency is determined, the water-holding capacity of the watered surfaces is the only parameter missing to optimize water consumption and can be obtained with an independent measurement.

## 2. Materials and Methods

Street surface temperatures were investigated on rue du Louvre, near Les Halles in the 1$^{st}$ and 2$^{nd}$ Arrondissements in Paris, France over the summer of 2013. Measurements were collected during the same experiment as that described by Hendel et al. [3]. Watered and control weather station positions are illustrated in Figure 1, however only data from the watered site will be studied. Both watered and dry portions of the street are approximately 180 m long and 20 m wide. Rue du Louvre has an aspect ratio approximately equal to one and has a N-NE–S-SW orientation. Data is presented in local daylight savings time (UTC +2).

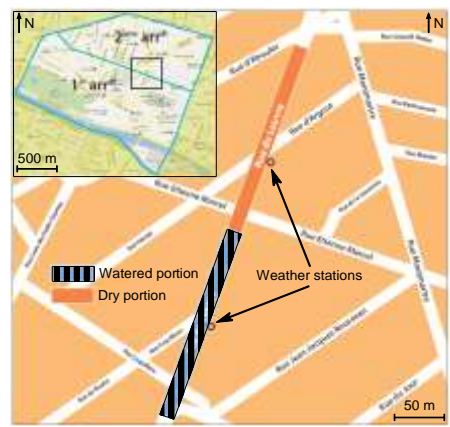

Figure 1: *Map of the rue du Louvre site*

## 2.1. Instruments

Surface temperature measurements were made by a FLIR B400 thermal camera with a spectral range of 7.5-13 µm. The camera was placed on the roof terrace of 46, rue du Louvre, approximately 20 meters above street level. The camera recorded false-color I.R. thermal and visible images simultaneously every hour on non-watered days and every 10 or 15 minutes on watered days. These images were used to estimate pavement surface temperature.

Apparent (measured) surface temperatures, also known as brightness temperatures, were corrected with the parameters indicated in Table 1, as measured by a weather station installed in front of 46, rue du Louvre, in order to obtain directional radiometric temperatures. The emissivity of the studied surfaces was obtained by the reference black body method, using an adhesive with a known emissivity of 0.95. Mean radiant temperature (MRT), calculated from the weather station measurements, was used as the reflected temperature for I.R. camera measurement corrections. Although this causes an underestimation of directional infrared temperatures since MRT is greater than the actual sky irradiance temperature, the error remains limited due to the high emissivity of the studied surfaces.

| | |
|---|---|
| Emissivity | 0.97 |
| Distance to target (height) | 20 m |
| Reflected temperature | Mean radiant temperature (MRT) measured by weather station |
| Relative humidity | As measured at 1.5 m by weather station |
| Air temperature | As measured at 1.5 m by weather station |

Table 1 : *Parameters used to correct apparent surface temperature*

Table 2 describes the instruments used for the purpose of this analysis. The weather station instruments within pedestrian reach were protected behind a cylindrical white-painted steel cage. Black globe temperature, air temperature and wind speed measurements from the weather station were used to estimate MRT using the method described by ASHRAE [15]. Errors introduced by the use of 4-m rather than 1.5-m wind speed in the calculation of MRT are neglected, as well as the influence of the steel cage.

| Parameter | Instrument | Height | Accuracy |
|---|---|---|---|
| Surface temperature | FLIR B400 infrared camera (7.5 – 13 µm) | Surface | 2°C |
| Relative humidity | Capacitive hygrometer | 1.5 m | 1.5% RH |
| Air temperature | Pt100 1/3 DIN B | 1.5 m | 0.1°C |
| Black globe temperature | Pt100 1/2 DIN A ISO 7726 | 1.5 m | 0.15°C |
| Wind speed | 2D ultrasonic anemometer | 4 m | 2% |

Table 2: *Instrument type, measurement height and accuracy*

The camera operated continuously from 08:00 on July 8[th] until 08:00 on September 6[th]. Interruptions occurred between 19:20 on July 8[th] until 18:00 on July 11[th] and from 18:00 on July 12[th] until 14:30 on July 15[th]. These are due to failures of the camera's time-lapse computer.

The surface temperatures of three street areas were surveyed as illustrated in Figure 2 (left): pavement zone 1, located above the pavement sensor used in the parent analysis [3], pavement zone 2, located further towards the street's center and a sidewalk zone. All three had an emissivity of 0.97 despite different surface composition and texture. No change in emissivity was applied for the wet pavement. It is assumed that the wet pavement has the emissivity of a water film, i.e. 0.98.

The studied zones were selected according to the presence of thermal disturbance sources, detected via a nighttime infrared photograph of the area provided in Figure 2 (right). The use of a nighttime rather than daytime infrared image prevents interpretation errors due to differences in insolation or shading. Thermally disturbed areas are either significantly warmer or colder than their surroundings, once differences in emissivity have been accounted for.

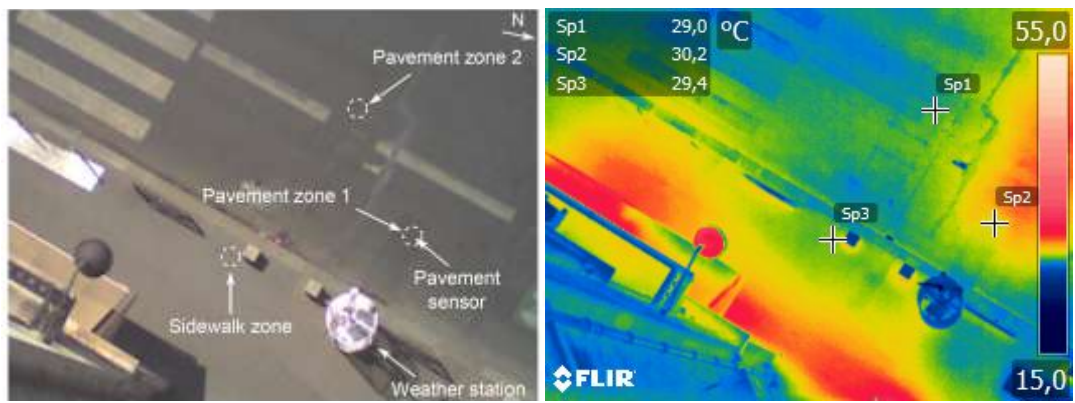

Figure 2: *Surface temperature measurement zones (left) and corresponding nighttime corrected infrared photograph on July 22$^{nd}$ at 03:20 (right). Temperature scale is in degrees Celsius. Sp2: pavement zone 1; Sp1: pavement zone 2; Sp3: sidewalk zone.*

The main disturbance is caused by a district heating main, which runs below the sidewalk along the edge of the building and crosses the street perpendicularly below the weather station. It is most easily recognized in its sidewalk portion in the form of a distinct red band running from left to right from the awning to the bottom edge of the false-color infrared image. It then runs perpendicularly, from bottom to top, underneath the weather station where it is not as visible and continues until the other side of the street, reappearing with rough contours just after pavement zone 1 (Sp2).

The presence of this main was confirmed by winter observations and the district heating operator. In the summertime, the heating network supplies minimal heat, solely for domestic hot water production. This heat source and the difference in overlying materials (see Table 3) are responsible for the temperatures of pavement zone 1 being up to 2°C warmer than the surrounding pavement at night. Pavement zone 2 was therefore added to complement the data collected in zone 1, above the pavement sensor. Another identified heat source is an air conditioning exhaust vent emitting hot air underneath the awning of the nearby grocery shop on the left side of the image in front of the pedestrian crosswalk. This exhaust vent explains the hot spot visible at the center of the awning itself and on the sidewalk directly below it. This heat source does not significantly affect the weather station measurements due to its position and height, however the sidewalk zone was selected to avoid its area of influence as well as that of the district heating main and areas shaded by other obstacles.

Table 3 describes the underlying materials and their respective thicknesses for each zone. Only the composition of pavement zone 1 is known with certainty thanks to the construction work undertaken to place the pavement sensor 5 cm below the surface. The other pavement structures are described based on data provided by the Roads and Traffic Division. It should be noted that the sidewalk's asphalt surface course does not include medium or coarse aggregates such as those used in asphalt concrete and is therefore much smoother than the road pavement surfaces. Furthermore, the odd road composition below pavement zone 1 is most likely attributable to the presence of the district heating main.

| Zone | Layer | Composition | Thickness |
| --- | --- | --- | --- |
| Pavement zone 1 | Surface | Cold-mixed asphalt concrete | 6 cm |
| | Base | Concrete | 34 cm |
| | Subgrade | Compacted ground | - |
| Pavement zone 2 | Surface and Binder | Hot-mixed asphalt concrete | 16 cm |
| | Base | Cement-treated base material | 20 cm |
| | Subgrade | Compacted ground | - |
| Sidewalk zone | Surface | Asphalt | 2 cm |
| | Base | Concrete | 10 cm |
| | Subgrade | Compacted ground | - |

Table 3: *Pavement structure in each zone*

## 2.2. Watering method

Watering was conducted if certain weather conditions were met according to Météo-France's three-day forecast. These as well as those for heat-wave warnings are presented in Table 4. The weather conditions required for watering were purposely set at a lower level than heat-wave warnings as these remain rare events. With this threshold, a total of 10 days were watered over the summer (July 8th, 9th, 10th, 16th, 22nd and 23rd (morning only), August 1st, 2nd and 23rd and September 5th 2013).

| Parameter | Watering starts | Heat-wave warning level |
| --- | --- | --- |
| Mean 3-day minimum air temperature ($BMI_{Min}$) | > 16°C | >21°C |
| Mean 3-day maximum air temperature ($BMI_{Max}$) | > 25°C | >31°C |
| Wind speed | < 10 km/h | - |
| Sky conditions | Sunny (less than 2 oktas cloud cover) | - |

Table 4: *Weather conditions for pavement-watering and heat-wave warnings*

Cleaning trucks assisted by manual operators sprinkled approximately 1 mm every hour from 06:30 to 11:30 and every 30 minutes from 14:00 until 18:30 on the sidewalk and pavement. This amount of water is considered to be the maximum water-holding capacity of the road surface in Paris by the Roads and Traffic Division. Exact watering times were reported by truck drivers, cross-checked against visible images taken by the rooftop camera when available. Resulting watering time precision is estimated to be no better than 5 minutes. A picture of pavement-watering underway across the street from the studied street surfaces (the same pedestrian crosswalk is visible) is presented in Figure 3. Each watering cycle required two passes by the cleaning truck to water both halves of the street.

Water used for this experiment was supplied by the city's 1,600 km non-potable water network, principally sourced from the Ourcq Canal. This water network is currently regarded by urban managers as having high sustainability potential for urban uses that do not require potable water such as green space irrigation [16].

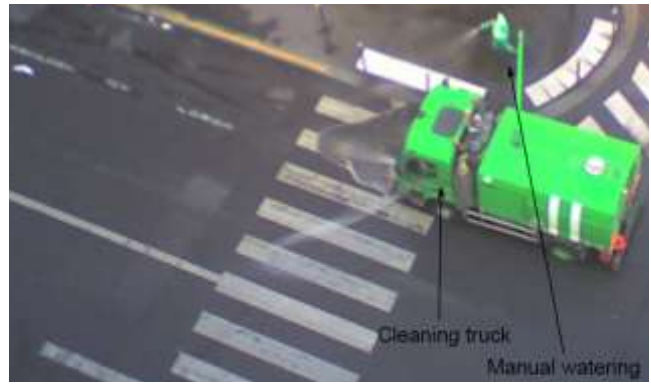

Figure 3: *Pavement- and sidewalk-watering on July 8$^{th}$ by a cleaning truck and its operators. Watering is taking place across the street from the weather station visible in Figure 2.*

## 2.3. Data analysis

For better comparability between dry and watered days, temperature data is presented over 24-hour periods extending from 06:00 until 05:59 on the next day, i.e. July 20$^{th}$ refers to data from 06:00 on July 20$^{th}$ until 05:59 on July 21$^{st}$.

I.R. camera measurements from July 8$^{th}$, 22$^{nd}$ (watered), 20$^{th}$ and 21$^{st}$ (control) will now be considered in the following section. Unfortunately, it is not possible to compare wet pavement surface temperatures from July 8$^{th}$ with dry pavement surface temperatures made a few days before or later in comparable weather conditions. The closest days were July 7$^{th}$ or 11$^{th}$ which had clear skies (less than 2 oktas cloud cover) and low wind speeds (less than 3 m/s), i.e. of Pasquill Atmospheric Stability Class A-B or more [17]. July 20$^{th}$ is the closest control day with available data which meets these weather criteria, however atmospheric temperatures were warmer than on July 8$^{th}$, with a respective BMI$_{Min}$ and BMI$_{Max}$ of 19° and 30.3°C on July 20$^{th}$ versus 18° and 29°C on July 8$^{th}$.

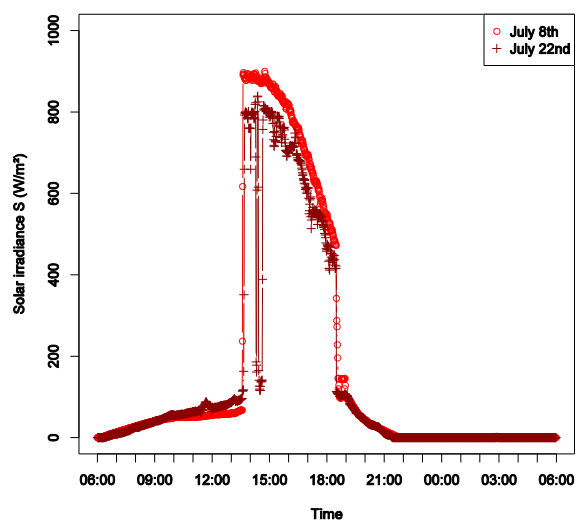

Figure 4: *Solar irradiance measured on July 8$^{th}$ and 22$^{nd}$.*

Although sun trajectories can change very rapidly during certain periods of the year, these changes are relatively small during the month of July. The daylight period on July 20$^{th}$ is approximately 25 minutes shorter than on July 8[th], while solar zenith is reduced by less than 2°. The magnitude of these changes can be seen in Figure 4 which illustrates solar irradiance as measured by the weather station on July 8[th] and 22[nd] for comparison. Daily shading conditions were checked and were found to be unmodified for the studied surfaces between July 8[th] and July 21[st]. For the purpose of this discussion, we therefore consider the changes in sun trajectory to be negligible.

Due to the different positions, pavement zone 1 and 2 and the sidewalk zone are not insolated simultaneously. Pavement zone 2 becomes insolated first at approximately 13:15, followed by pavement zone 1 at 13:35 and finally the sidewalk zone at 13:55.

## 3. Street Surface Temperatures

I.R. camera measurements from July 8[th], 22[nd] (watered), 20[th] and 21[st] (control) will now be considered in order to study the effects of pavement-watering on surface temperatures. The results of this analysis will be used to formulate recommendations to improve the effectiveness of the watering method.

### 3.1. Results

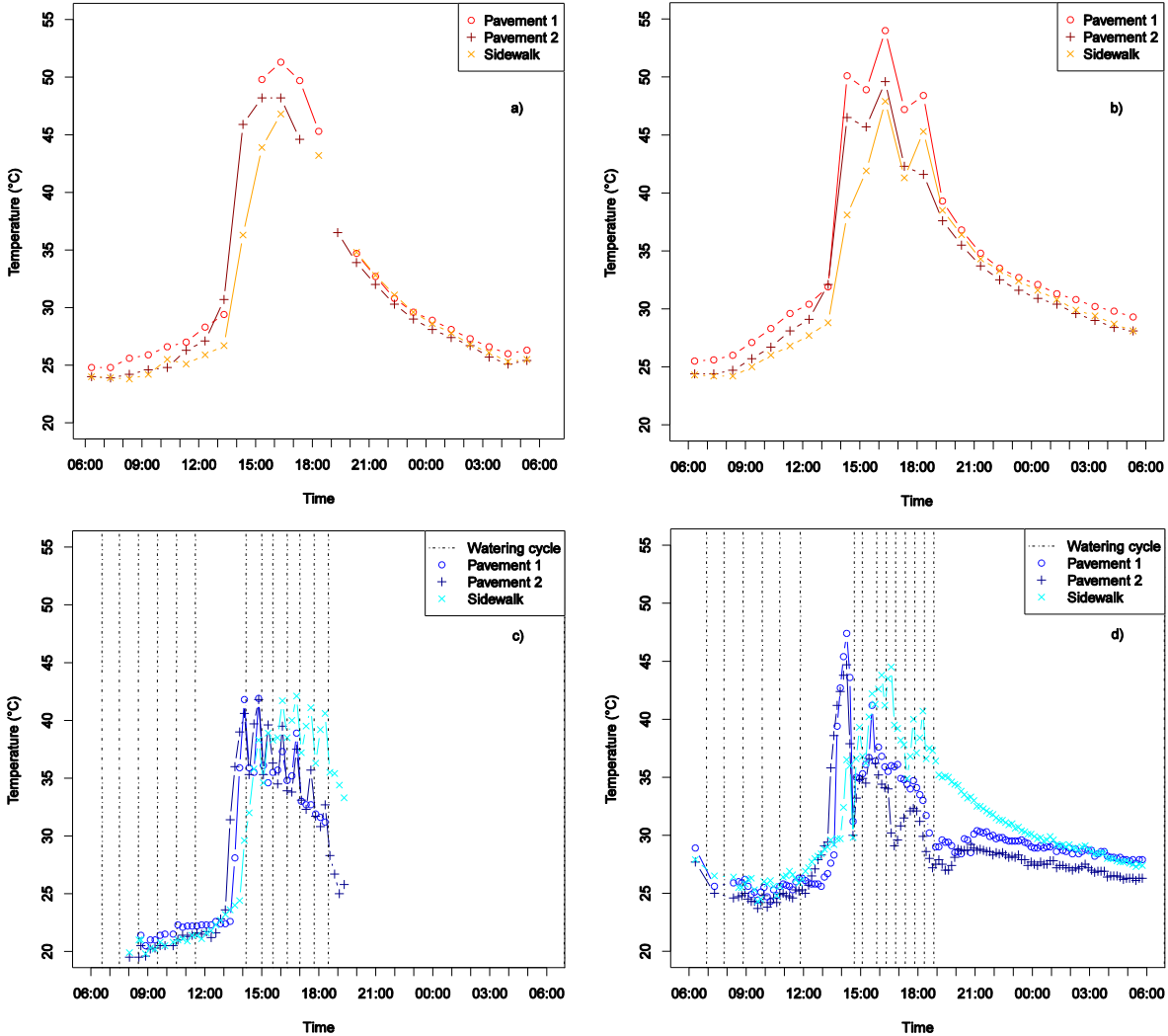

Figure 5: *Pavement Directional Radiometric Temperature on control days: a) July 20[th] and b) July 21[st]; and on watered days: c) July 8[th] and d) July 22[nd]*

Figure 5 presents directional radiometric temperature for the three defined areas of the street: the (o) series represents pavement zone 1; the (+) series represents pavement zone 2; the (x) series represents the sidewalk. Some data points are missing due to passing or parked vehicles over the measurement areas. Vertical dot-dashed lines indicate watering cycles for July 8$^{th}$ and 22$^{nd}$. It should be noted that afternoon watering on July 8$^{th}$ occurred about every 45 minutes, while it was every 30 minutes on July 22$^{nd}$.

On July 20$^{th}$ and 21$^{st}$, surface temperatures reach a low of 24°C between 06:00 and 07:00 and a high of 50-54°C at 16:00. All three surfaces follow a very similar approximately bell-shaped curve. Morning temperatures increase slowly until they spike when the pavement becomes insolated around 13:30. Pavement shading causes an exponentially-shaped decrease in temperatures starting between 18:00 and 19:00. Pavement zone 1 is generally the warmest, followed by pavement zone 2 and the sidewalk, with nocturnal temperatures being the most similar with differences smaller than 2°C. Temperature differences are more pronounced during the day, particularly during insolation. It should be noted that the temperature fluctuations which occur on July 21$^{st}$ are due to the passing of isolated clouds.

On July 22$^{nd}$, the daily low was about 25°C, but this temperature was maintained from 07:30 until 12:00. A maximum temperature of about 47°C was reached just before afternoon pavement-watering began. The bell-shape of the temperature curves is greatly affected by watering. Morning increases are slower, insolated surface temperatures are several degrees lower and nighttime temperatures follow a more linear decrease, except for the sidewalk zone. After watering begins, the temperature of all surfaces remains below 45°C. Furthermore, unlike on control days, the sidewalk becomes the warmest area after afternoon watering begins and remains so until late in the night. Similar temperatures and trends are observed in available data from July 8$^{th}$.

A dip in the temperature of pavement zone 2 and that of the sidewalk should be noted on July 22$^{nd}$. It occurs due to shade caused by the buildings across the street at approximately 17:00 and 18:00, respectively. This shade event did not occur before this date.

In addition to these changes, the presence of several temperature spikes during and just before afternoon watering is noted on both July 8$^{th}$ and 22$^{nd}$. For pavement zones 1 and 2, these are mainly found on July 8$^{th}$, with nearly all occurring during afternoon watering. On July 8$^{th}$, the detected spikes are up to two 15-minute measurements wide. On July 22$^{nd}$, when watering occurs more frequently, only two spikes are visible, excluding the temperature dip due to building shading: one before and one during afternoon watering. The latter spike follows a 50-minute interruption in pavement-watering and is only detected by one measurement. For the sidewalk, spikes are found on both watered days. The amplitude of these local maxima is in the order of 5°C on July 8$^{th}$ for all three zones. On July 22$^{nd}$, the amplitude of the sidewalk maxima is reduced to about 2-3°C while temperatures are comparable. In the morning of July 22$^{nd}$, small temperature spikes may occur for the sidewalk zone, but these are of low amplitude and are hard to distinguish.

Spikes just before afternoon watering begins occur for the pavement zones only. This is because the studied areas are not insolated simultaneously and pavement-watering began before the sidewalk was insolated, but not before the pavement zones were.

Table 5 summarizes the average changes in surface temperatures between 06:30 and 13:00 and between 15:00 and 18:30 on July 22$^{nd}$ (watered) compared to the average observations from July 20$^{th}$ and 21$^{st}$ (control).

Average temperature reductions on July 22$^{nd}$ are most important in the afternoon and most pronounced for pavement zone 1 followed by pavement zone 2. The sidewalk has the lowest average cooling of the three zones. In fact, the sidewalk is actually found to be 1.1°C warmer on the morning of July 22$^{nd}$ than on July 20$^{th}$ or 21$^{st}$, despite pavement-watering. This is attributed to temperatures being about 5°C warmer in the hours preceding 06:00 on July 22$^{nd}$ compared to July 20$^{th}$ and 21$^{st}$ (see Figures 5 a), b) and d)). While radiometric temperatures were about 25°C at 06:00 on July 20$^{th}$ and between 03:00 and 06:00 on July 21$^{st}$, reduced nighttime cooling, due to warmer meteorological conditions, on July 21$^{st}$ caused these temperatures to increase to roughly 30°C. Therefore, the calculation of the morning cooling effect on July 22$^{nd}$ should take this aspect into account. When this is the case, pavement-watering is found to reduce morning temperatures by about 4°C for the pavement surfaces and 2°C for the sidewalk.

|               | Pavement 1 (o) | Pavement 2 (+) | Sidewalk (x) |
|---------------|----------------|----------------|--------------|
| 06:30 – 13:00 | 1.6°C          | 0.73°C         | -1.1°C       |
| 15:00 – 18:30 | 13°C           | 11°C           | 6.2°C        |

Table 5: *Average temperature reductions observed on July 22$^{nd}$ (watered) compared to average control day temperatures (July 20$^{th}$ and 21$^{st}$)*

### 3.2. Discussion

Pavement-watering was found to reduce morning temperatures between 2° and 4°C and between 6° and 13°C for afternoon surface temperatures, depending on the surveyed area. Also, the daily minimum temperature on watered days was found to be the same as on control days despite warmer temperatures the night before. In addition, pavement-watering maintained this minimum temperature for several hours instead of it being reached only briefly on control days. The maximum temperature reached was also reduced by several degrees. Finally, temperature spikes were observed on watered days, both before and during afternoon pavement-watering, which were not present on control days. No clear unquestionable spikes were observed in the morning.

These effects will help limit the conditions favorable to the formation of urban heat islands without compromising pedestrian comfort by simultaneously reducing atmospheric sensible and radiant heat fluxes. Indeed, while high albedo pavements also have lower surface temperatures and thus reduce sensible heat flows to the atmosphere, they have been shown to increase daytime pedestrian radiant loads via increased shortwave reflection [18]. In the case of pavement-watering, surface wetting will tend to decrease pavement albedo, thus lowering reflected shortwave radiation (for further details on albedo change due to surface wetting, see Lekner & Dorf [19]). However, the expected increase in air humidity caused by watering may negatively affect pedestrian comfort.

Differences in surface temperatures in zones 1 and 2 may be explained by differences in pavement materials (hot- versus cold-mixed asphalt concrete), pavement geometry (different surface slopes) and the presence of the district heating main.

These pavement temperature observations are consistent with those made by Kinouchi & Kanda [5], [6] and Yamagata et al. [8] who also conducted pavement-watering experiments in urban environments in Japan with regular or pervious asphalt pavements in similar conditions.

The local temperature spikes found on watered days, both before afternoon watering begins and in between cycles afterwards, allow us to make recommendations to improve the effect of pavement-watering: afternoon watering should begin just before surface insolation to limit the initial temperature increase, while a watering cycle period of 30 minutes for the pavement seems sufficient to erase spikes which occur during afternoon pavement-watering.

However, a 30-minute watering interval remains too long for the sidewalk which requires a higher frequency. One-hour watering is sufficient for morning shaded conditions for all surfaces.

Indeed, the first spikes in surface temperatures are reached because the first afternoon watering cycle occurs too late compared to the last morning watering cycle. With more than two hours since the last watering cycle, the studied surface areas were dry when they became exposed to direct sunlight. Had watering begun a few minutes before insolation, the surfaces would have been wet when they became insolated. The water would then have limited the initial temperature increase and lowered the maximum surface temperature reached. This is also the case of the sidewalk, however no maxima are visible before afternoon watering began since it remained shaded until watering resumed.

The spikes reached once afternoon watering is underway indicate an insufficient watering frequency. They occur in between watering cycles, only if the surface has had enough time to dry out, as is confirmed by corresponding visible images (not shown). Increasing the watering frequency should therefore erase these temperature spikes. This is confirmed when comparing pavement temperatures from July $8^{th}$ and $22^{nd}$, when the watering cycles occurred every 45 and 30 minutes, respectively. While 45 minutes was too long, 30 minutes was sufficient for the pavement. Since spikes are still visible even with 30-minute watering for the sidewalk, a higher watering frequency is necessary to reach maximum cooling. This is due to the sidewalk's asphalt surface course whose texture is significantly smoother than that of the pavement zones. It therefore retains less water than the pavement, resulting in shorter cooling due to increased runoff. The difference in watering method, with the manual hose rather than the vehicle's sprinkler, may also explain part of this difference as it was more difficult to uniformly water the sidewalk due to the presence of pedestrians and other obstacles.

As no temperature spikes are visible in the morning, the one-hour watering frequency is deemed sufficient.

From this analysis, only the water-retaining capacities of the studied surfaces are needed to simultaneously optimize the method's water consumption.

However, because the measurement frequency is 10 minutes at best, surface temperature spikes may still exist in the last minutes prior to watering cycles that are not detected by the I.R. camera. Closer analysis of the temporal distribution of the data shows that measurements occur as little as 4 minutes before watering cycles on July $22^{nd}$, without the detection of spikes. However, this does not fully exclude the possibility of undectected spikes during these last 4 minutes.

To remedy this, measurements can either be more frequent or they can be synchronized with watering cycles so as to occur a few seconds before watering. In the case of a manual watering method such as ours, it is not possible to reliably synchronize measurements with watering cycles. Watering cycles are not distributed as regularly as planned due to disturbances such as human error, traffic or the interruption of watering due to the presence of pedestrians. In this situation, the sampling frequency should be increased. In order to be sure that the error window is limited and no worse than what is described here, a 5-minute frequency is recommended at least.

In the case of an automated pavement-watering system, synchronisation should be more feasible and the sampling rate could even be reduced to the watering frequency: as long as the measurement is made right before watering, no spikes should be missed. However, if the watering frequency is deemed insufficient following the data analysis, this sampling rate provides no indication on what the correct frequency might be, thus requiring multiple follow-up field trials. A frequency at least two or three times as high as the watering frequency will help narrow down the optimal watering frequency band significantly, saving time and effort.

When the surface temperature data is compared to Hendel et al.'s pavement heat flux data recorded 5 cm below the pavement surface, good agreement of the observations and the recommendations is found [3]. This is due to good correspondence between heat flux and surface temperature spike observations. It should be noted however that temperature spikes have a relatively greater amplitude than heat flux spikes. This is unsurprising as heat flux 5 cm below the surface is naturally less sensitive to surface conditions than surface temperatures are.

However, this greater sensitivity means that minor heat flux spikes that may have been disregarded, such as those apparent in the afternoon of July $10^{th}$ [3], may correspond to significant surface temperature spikes. The absence of surface temperature data on that date prevents us from exploring this possibility further, but these considerations emphasize the need for measurements within the last few minutes preceding watering cycles.

As a result, it is tempting to increase the frequency further. 1-minute measurements would be ideal and would ensure that no significant spikes are missed. The amount of time required to empirically determine the optimal watering frequency would be significantly reduced as well since only one trial would be needed if no field-of-view obstructions occur.

Such a high frequency creates large numbers of data files. Apart from requiring a higher storage capacity, the treatment load will also increase significantly if corrections cannot be integrated automatically and must be applied manuallly. For this reason, automatic corrections are necessary for 1-minute measurements to be simple to analyze and interpret.

However, given that the information sought after is the detection of temperature spikes, the corrections may be unnecessary in certain situations. Indeed, if the I.R. imaging device is only a few meters away from its target and the surface material has high emissivity, i.e. both atmospheric and reflected temperature influence are low, uncorrected brightness temperature should suffice to detect surface drying in the form of temperature spikes. In this case, there is no reason not to choose 1-minute measurements if storage capacity is sufficient.

To summarize, if the instrument data is readily useable for analysis, either because brightness temperature is sufficient or corrections are automatically applied, 1-minute measurements are recommended. Otherwise, if manual data correction and analysis are too resource-intensive, 5-minute measurements should be enough, especially if these can be synchronised with the watering cycles.

## 4. Conclusion

The field study conducted on rue du Louvre in Paris over the summer of 2013 has allowed us to explore the effects of pavement-watering on street surface temperatures. They were found to be reduced by several degrees, both during watering and several hours after it had ended. These effects will tend to reduce the amplitude of the sensible heat flux absorbed by the atmosphere, thus contributing to reduce the conditions favorable to the formation of urban heat islands without increasing radiant loads received by pedestrians. Expected increases in humidity however may negatively affect pedestrian comfort, although they should occur during watering when conditions are quite dry (less than 40% relative humidity).

In addition, surface temperature measurements were found to provide useful information for optimizing the watering method. Based on the 10-minute measurements made here, temperature spikes were detected. Spikes occurring before pavement-watering resumed in the afternoon indicated that watering should begin before surface insolation. Spikes were also found in between watering cycles, indicating that the watering frequency should be increased to reduce surface drying. For asphalt concrete pavement surfaces in a N-S street in Paris in the month of July, watering every 30 minutes during insolation and every hour during shading is sufficient. This holds true for asphalt sidewalks during shading, however 30 minutes was found to be too long during insolation. These conclusions agree well with previous work by the authors based on pavement heat flux and solar irradiance measurements.

The pavement-watering frequency can therefore be successfully optimized on the basis of the detection of surface drying, manifested in the form of rapid surface temperature changes. Time-lapse infrared camera temperature surveys are able to meet this requirement if they are frequent enough to detect the temperature spikes which may occur in the last moments preceding watering cycles. A measurement rate of 5 minutes or less ensures that minimal temperature spikes are omitted, while 1-minute measurements are ideal where possible.

Pyrometers, which measure spot surface temperatures and are significantly cheaper, could also be used, but unlike I.R. cameras they do not allow the simultaneous survey of different areas. Given their lower cost, it may still be more feasible to install several of these in order to compensate for this weakness. Installation of either instrument in the street is simple and non-invasive compared to pavement sensors, making them well suited to dense urban environments. Street lamps make suitable candidates for power supply requirements, but many Parisian streets have lamps mounted directly onto building façades where adding additional weight is difficult. Furthermore, since vehicles, pedestrians and other objects may enter the line-of-site of radiometric instruments, it is recommended to include a synchronized webcam. As was seen in the case of our site, many thermal disturbance sources exist in urban environments and need to be clearly identified before instrument targets are selected in order to avoid misinterpretation. I.R. photographs can be of valuable assistance in this task.

Furthermore, if this method is combined with an independent measurement of the water-holding capacity of the pavement, the method's total water consumption and watering frequency can be optimized simultaneously.

The described method therefore provides a feasible alternative to pavement heat flux and solar measurements for officials wishing to test several characteristic streets in order to devise a city-wide watering strategy.

**Acknowledgements**

The authors would like to thank the Orange Group for allowing the use of their rooftop terrace located at 46, rue du Louvre for instruments used during this experiment. They also acknowledge the support of Météo-France and APUR as well as the Green Spaces and Environment, Roads and Traffic and the Waste and Water Divisions of the City of Paris during the preparation phase of this experiment.

Funding for this experiment was provided for by the Water and Sanitation Department of the City of Paris.

**Nomenclature**

*APUR*  Parisian urban planning agency
$BMI_{Min}$  Minimum biometeorological index, 3-day mean of daily minimum temperature, °C
$BMI_{Max}$  Maximum biometeorological index, 3-day mean of daily maximum temperature, °C
*MRT*  mean radiant temperature, °C
*UHI*  urban heat island

# References

[1] Météo-France and CSTB, "EPICEA Project - Final Report," Paris, France (in French), 2012.

[2] M. Bouvier, A. Brunner, and F. Aimé, "Nighttime watering streets and induced effects on the surrounding refreshment in case of hot weather. The city of Paris experimentations," *Tech. Sci. Méthodes*, no. 12, pp. 43–55 (in French), 2013.

[3] M. Hendel, M. Colombert, Y. Diab, and L. Royon, "An analysis of pavement heat flux to optimize the water efficiency of a pavement-watering method," *Appl. Therm. Eng.*, no. (under review), 2014.

[4] P. Maillard, F. David, M. Dechesne, J.-B. Bailly, and E. Lesueur, "Characterization of the Urban Heat Island and evaluation of a road humidification mitigation solution in the district of La Part-Dieu, Lyon (France)," *Tech. Sci. Méthodes*, no. 6, pp. 23–35 (in French), 2014.

[5] T. Kinouchi and M. Kanda, "An Observation on the Climatic Effect of Watering on Paved Roads," *J. Hydrosci. Hydraul. Eng.*, vol. 15, no. 1, pp. 55–64, 1997.

[6] T. Kinouchi and M. Kanda, "Cooling Effect of Watering on Paved Road and Retention in Porous Pavement," in *Second Symposium on Urban Environment*, 1998, pp. 255–258.

[7] R. Takahashi, A. Asakura, K. Koike, S. Himeno, and S. Fujita, "Using Snow Melting Pipes to Verify the Water Sprinkling's Effect over a Wide Area," in *NOVATECH 2010*, 2010, p. 10.

[8] H. Yamagata, M. Nasu, M. Yoshizawa, A. Miyamoto, and M. Minamiyama, "Heat island mitigation using water retentive pavement sprinkled with reclaimed wastewater," *Water Sci. Technol. a J. Int. Assoc. Water Pollut. Res.*, vol. 57, no. 5, pp. 763–771, Jan. 2008.

[9] T. Nakayama and T. Fujita, "Cooling effect of water-holding pavements made of new materials on water and heat budgets in urban areas," *Landsc. Urban Plan.*, vol. 96, no. 2, pp. 57–67, May 2010.

[10] T. Nakayama, S. Hashimoto, and H. Hamano, "Multiscaled analysis of hydrothermal dynamics in Japanese megalopolis by using integrated approach," *Hydrol. Process.*, vol. 26, no. 16, pp. 2431–2444, Jul. 2012.

[11] A. Lemonsu, R. Kounkou-Arnaud, J. Desplat, J.-L. Salagnac, and V. Masson, "Evolution of the Parisian urban climate under a global changing climate," *Clim. Change*, vol. 116, no. 3–4, pp. 679–692, Jul. 2012.

[12] J.-M. Robine, S. L. K. Cheung, S. Le Roy, H. Van Oyen, C. Griffiths, J.-P. Michel, and F. R. Herrmann, "Death toll exceeded 70,000 in Europe during the summer of 2003.," *C. R. Biol.*, vol. 331, no. 2, pp. 171–178, Feb. 2008.

[13] D. Li and E. Bou-Zeid, "Synergistic Interactions between Urban Heat Islands and Heat Waves: The Impact in Cities Is Larger than the Sum of Its Parts," *J. Appl. Meteorol. Climatol.*, vol. 52, no. 9, pp. 2051–2064, Sep. 2013.

[14] A. Burton, H. J. Fowler, S. Blenkinsop, and C. G. Kilsby, "Downscaling transient climate change using a Neyman–Scott Rectangular Pulses stochastic rainfall model," *J. Hydrol.*, vol. 381, no. 1–2, pp. 18–32, Feb. 2010.

[15] ASHRAE, *ASHRAE Fundamentals Handbook 2001*, SI Edition. American Society of Heating, Refrigerating, and Air-Conditioning Engineers, 2001.

[16] APUR, "Study on the future of the non-potable water network - Part 1: Diagnostic and analysis," Paris, France (in French), 2010.

[17] F. Pasquill, "The estimation of the dispersion of windborne material," *Meteorol. Mag.*, vol. 90, no. 1063, pp. 33–49, 1961.

[18] E. Erell, D. Pearlmutter, D. Boneh, and P. B. Kutiel, "Effect of high-albedo materials on pedestrian heat stress in urban street canyons," *Urban Clim.*, Oct. 2013.

[19] J. Lekner and M. C. Dorf, "Why some things are darker when wet.," *Appl. Opt.*, vol. 27, no. 7, pp. 1278–80, Apr. 1988.